\documentclass[journal=tches]{iacrtrans}

\usepackage{graphicx}
\usepackage{xcolor}
\usepackage{pifont}
\usepackage{xspace}
\usepackage{url}
\usepackage[mode=buildnew]{standalone}
\usepackage{tikz}
\usepackage{booktabs}
\usepackage{multirow}
\usetikzlibrary{positioning, trees, arrows, calc, automata}
\usepackage{hyperref}
\usepackage{cleveref}

\definecolor{col1}{HTML}{4398d1}
\definecolor{col2}{HTML}{ff4842}
\definecolor{col3}{RGB}{ 38, 93,105}
\definecolor{col4}{RGB}{ 44,127, 66}

\definecolor{greenc}{HTML}{64c37d}
\definecolor{redc}{HTML}{e13957}

\ifstandalone
    \newcommand{\icon}[1]{../icons/#1}
\else
    \newcommand{\icon}[1]{images/icons/#1}
\fi

\newcommand{\imgmemory}{\includegraphics[width=2.0cm]{\icon{computerpack/013-ram}}}
\newcommand{\imggpu}{\includegraphics[width=1.4cm]{\icon{computerpack/002-vga}}}
\newcommand{\imggpusmall}{\includegraphics[width=1.0cm]{\icon{computerpack/002-vga}}}
\newcommand{\imgmouse}{\includegraphics[width=1.4cm]{\icon{computerpack/017-mouse}}}
\newcommand{\imgkeyboard}{\includegraphics[width=1.4cm]{\icon{computerpack/024-keyboard}}}

\newcommand{\imglan}{\includegraphics[width=1.4cm]{\icon{computerpack/023-lan}}}

\newcommand{\imgdisplay}{\includegraphics[height=1.4cm]{\icon{computerpack/021-mobile}}}

\newcommand{\imgsim}{\includegraphics[height=1.4cm]{\icon{computerpack/008-sim-card}}}

\newcommand{\imgsd}{\includegraphics[height=1.4cm]{\icon{computerpack/011-sd-card}}}

\newcommand{\imgcamera}{\includegraphics[width=1.4cm]{\icon{computerpack/035-camera}}}

\newcommand{\imgenclave}{\includegraphics[width=2.0cm]{\icon{enclave}}}

\newcommand{\imgenclavesmaller}{\includegraphics[width=1.0cm]{\icon{enclave}}}

\newcommand{\imgenenclaveredsmall}{\includegraphics[width=1.0cm]{\icon{sgx_red}}}

\newcommand{\imglocklarge}{\includegraphics[width=0.5cm]{\icon{lock-icon}}}

\let\oldding\ding%
\renewcommand{\ding}[2][1]{\scalebox{#1}{\oldding{#2}}}%

\newcommand{\one}{\ding[1.2]{172}\xspace}
\newcommand{\two}{\ding[1.2]{173}\xspace}
\newcommand{\three}{\ding[1.2]{174}\xspace}

\newcommand{\ce}{driver enclave\xspace}

\newcommand{\Ce}{Driver enclave\xspace}

\newcommand{\app}{application enclave\xspace}
\newcommand{\App}{Application enclave\xspace}

\newif\ifremoveall{}
\removealltrue{}

\ifremoveall{}
\newcommand{\aritra}[1]{\textbf{\emph{ #1 \colorbox{green}{[Aritra]}}}}
\newcommand{\ip}[1]{\textbf{\emph{ #1 \colorbox{cyan}{[Ivan]}}}}
\newcommand{\moritz}[1]{\textbf{\emph{ #1 \colorbox{yellow}{[Moritz]}}}}

\newcommand{\todo}[1]{\textcolor{red}{TODO: \@#1}}
\newcommand{\todoref}{\textcolor{red}{[ref]}}
\newcommand{\citneed}{[\textcolor{red}{cit}] }
\else
\newcommand{\aritra}[1]{}
\newcommand{\ip}[1]{}
\newcommand{\moritz}[1]{}
\newcommand{\todo}[1]{}
\newcommand{\todoref}[1]{}
\newcommand{\citneed}[1]{}
\fi

\newcommand{\name}{composite enclave\xspace}

\def\nameenclave{unit enclave\xspace}

\def\Nameenclave{Unit enclave\xspace}
\def\NAMEENCLAVE{Unit Enclave\xspace}

\newcommand{\sphw}[0]{specialized hardware\xspace}
\newcommand{\Sphw}[0]{Specialized hardware\xspace}
\newcommand{\SPHW}[0]{Specialized Hardware\xspace}

\newcommand{\myparagraph}[1]{\paragraph{#1}}

\newcounter{para}
\newcommand{\mypara}[1]{\refstepcounter{para}\paragraph{\thepara.\space{}#1}}

\newcounter{myctr}

\expandafter\def\expandafter\UrlBreaks\expandafter{\UrlBreaks
  \do\a\do\b\do\c\do\d\do\e\do\f\do\g\do\h\do\i\do\j
  \do\k\do\l\do\m\do\n\do\o\do\p\do\q\do\r\do\s\do\t
  \do\u\do\v\do\w\do\x\do\y\do\z\do\A\do\B\do\C\do\D
  \do\E\do\F\do\G\do\H\do\I\do\J\do\K\do\L\do\M\do\N
  \do\O\do\P\do\Q\do\R\do\S\do\T\do\U\do\V\do\W\do\X
  \do\Y\do\Z}

\graphicspath{{images/}}

\title{Composite Enclaves: Towards Disaggregated Trusted Execution}

\author{Moritz Schneider\thanks{Equal contribution}\inst{1} \and Aritra Dhar\textsuperscript{*}\thanks{Work was done while at ETH Zurich}\inst{2} \and Ivan Puddu\inst{1} \and Kari Kostiainen\inst{1} \and Srdjan \v{C}apkun\inst{1}}
\authorrunning{Moritz Schneider, Aritra Dhar, Ivan Puddu, Kari Kostiainen, Srdjan \v{C}apkun}

\institute{ETH Zurich, Zurich, Switzerland, \email{firstname.lastname@inf.ethz.ch}
    \and Huawei Zurich Research Center, Switzerland, \email{firstname.lastname@huawei.com}}

\begin{document}
\maketitle

\keywords{Trusted execution environments \and RISC-V security}

\begin{abstract}
The ever-rising computation demand is forcing the move from the CPU to heterogeneous specialized hardware, which is readily available across modern datacenters through disaggregated infrastructure. On the other hand, trusted execution environments (TEEs), one of the most promising recent developments in hardware security, can only protect code confined in the CPU, limiting TEEs' potential and applicability to a handful of applications. We observe that the TEEs' hardware trusted computing base (TCB) is fixed at design time, which in practice leads to using untrusted software to employ peripherals in TEEs. Based on this observation, we propose \emph{composite enclaves} with a configurable hardware and software TCB, allowing enclaves access to multiple computing and IO resources. 
Finally, we present two case studies of composite enclaves: i) an FPGA platform based on RISC-V Keystone connected to emulated peripherals and sensors, and ii) a large-scale accelerator. These case studies showcase a flexible but small TCB (2.5 KLoC for IO peripherals and drivers), with a low-performance overhead (only around 220 additional cycles for a context switch), thus demonstrating the feasibility of our approach and showing that it can work with a wide range of specialized hardware.

\end{abstract}

\section{Introduction}
\label{sec: intro}

For most of the computer's history, designing an architecture around the CPU allowed extracting the most performance benefits from Moore's law.  %
Nowadays, however, the demand for increased computation power is usually met with special-purpose hardware: GPUs are often orders of magnitude more efficient than a CPU for parallel workloads such as graphics and machine learning, and FPGAs often achieve similar gains for custom workloads. Some tasks such as machine learning are even pervasive enough to justify the investment into fully custom ASICs~\cite{TPU}. %
In these modern platform architectures, the CPU's main job is to move data to relevant \sphw~\cite{spec_hw_acc}, collecting the results, and then possibly feeding them to yet another device. Effectively, the CPU's primary role is shifting towards a mere coordinator of available \sphw.
Cloud computing architectures are even adopting a disaggregated model called \emph{composable disaggregated infrastructure} (CDI)~\cite{disaggregatedcomp,lim2009disaggregated,nitu2018welcome} in which data centers no longer consist of a number of connected servers, but of functional blocks connected with high-speed interconnects. Each block provides a pool of a particular resource, be it GPUs, CPUs, memory, storage, or FPGAs, to allow for fine-grained resource allocation and acceleration. When more resources are requested, only a particular block needs to be augmented, rather than requiring the provisioning of full-fledged monolithic servers.

At the same time, the security of modern systems has also come under scrutiny due to numerous vulnerabilities related to the high complexity of operating systems and hypervisors~\cite{checkoway2013iago,suzaki2011memory}.
Because of this, it has become more attractive to rely on smaller and lower layers, i.e., firmware or even immutable hardware to enforce security and to reduce the underlying trusted computing base (TCB).
Most notably, this has led to the rise in trusted execution environments (TEEs). 
TEE designs vary to a large degree but, in general, they isolate execution environments without having to trust operating systems and hypervisors~\cite{costan2016intel,costan2016sanctum,winter2008trusted}. TEEs rely on hardware primitives of the CPU and only consider the CPU package to be trusted, while all the other hardware components of the platform are explicitly assumed malicious. 

These two developments present an apparent disconnect: on one side, modern computer architectures are increasingly distributing computation tasks onto (disaggregated) \sphw for performance and scalability. On the other, TEEs provide strong security guarantees for code and data that are confined within the CPU.
Combining \sphw and existing TEEs in the style of Intel SGX~\cite{costan2016intel} requires to trust the OS. E.g., the keyboard input to an SGX enclave can be read and altered by the untrusted OS. 
Another style of TEEs, notably ARM TrustZone~\cite{winter2008trusted}, allows to extend the hardware TCB at design time to on-chip peripherals. However, the secure OS~\cite{winter2008trusted} must include drivers to all \sphw devices even if they are not used for most enclaves. E.g., an enclave that needs user input through a keyboard also needs to trust a massive camera driver. The main reason for this shortcoming of existing TEEs lies in the statically assigned hardware TCB at design time by the CPU manufacturer.
End-users need to rely not only on a fixed hardware TCB, but also potentially end up including the (secure) OS into their software TCB. 
In short, current TEEs struggle to support \sphw while adhering to the principle of least privilege.

We propose a TEE with a configurable software and hardware TCB, a concept that we name \emph{composite  enclaves}. %
Composite enclaves, are formed by a collection of what we call \emph{\nameenclave{}s}, that are distributed over several hardware components. E.g., a \name{} can be composed of a \nameenclave on the GPU (or only some GPU cores) and one on the CPU. Like in traditional TEEs, a \name{} can be remotely attested. However, attestation to a \name{} not only reports a measurement of the software TCB but also of the hardware components that are part of the \name{}.

The shift towards configurable hardware and software TCBs has wide-ranging implications concerning integrity, confidentiality, and attestation of a \name. 
E.g., attestation to traditional TEEs allows verification of code integrity and the genuineness of the processor. However, attestation to a TEE with a flexible hardware TCB also requires verifying the composition of the hardware TCB. E.g., a remote verifier might want assurance that his enclave has exclusive access to a sensor. 
To account for a flexible hardware TCB, we extend the traditional attestation mechanism to include the system composition's integrity.
Moreover, the untrusted OS could remap specialized hardware devices at runtime with an untrustworthy device, which should not receive access to sensitive data. Therefore, \nameenclave{}s need to be informed upon any changes in the system's configuration and must be able to, e.g., halt execution until re-attestation. We call this property \emph{platform awareness} and achieve it by introducing two new events into the enclave life cycle, \textit{connect} and \textit{disconnect}, which allow tracking the liveliness of one \nameenclave{} from another.

We validate our design choices in a prototype (available online~\cite{ourcode}) that we develop on top of RISC-V and Keystone~\cite{keystone}. We make the key design decision to facilitate the communication between \sphw and the CPU with shared memory. This not only reduces the cost of context switches in enclave-to-enclave communication but also allows enclaves to communicate directly with \sphw, as these are memory-mapped, and allows to reuse existing drivers. Our prototype modifies the way Keystone uses RISC-V physical memory protection (PMP) to let enclave memory overlap, which enables shared memory. 
We perform an extensive security analysis of our prototype, analyzing the implications of our design with respect to side-channels, the \nameenclave{}'s interactions with peripherals, their life-cycles, and attestation. %
We further evaluate two case studies: first, we demonstrate an end-to-end prototype on an FPGA with simple peripherals emulated on a microcontroller; and second, we take an existing accelerator~\cite{zaruba2020manticore} and integrate it into a \name{}, adding support for multi-tenant isolation. In the first case study, we developed a prototype on top of an FPGA that is running a RISC-V core with keystone. The FPGA is connected to an Arduino microcontroller that emulates IO peripherals and sensors. It required around $2.5$ KLoC combined for the driver and the firmware changes to enable remote attestation. 
While the first case study focuses on IO peripherals and sensors that often requires exclusive access by an application, in the second case study, we demonstrate how to adapt an existing accelerator so it can support multi-tenant isolation and remote attestation. Here we enable multiple \name{}s to concurrently use the accelerator  while still giving meaningful isolation guarantees to remote verifiers. The TCB of Keystone increased by around $600$ lines of code (LoC) and the additional logic in the context switch increased by 220 cycles (from around 4700 to 4900 cycles).

In summary, the contributions of our paper are the following:

\begin{enumerate}
  \item We extend traditional TEEs with a configurable hardware TCB, i.e., the enclave's TCB only includes the driver, and firmware of the used \sphw. We call these new enclaves \emph{composite enclaves}. We identify two new properties that are relevant for these systems, a more comprehensive \emph{attestation} for \name{}s, and \emph{platform awareness}. Additionally, we propose a software design that abstracts the underlying hardware layer to ease the integration with the existing application and driver ecosystem. 
  
  \item We analyze the security aspects of our approach in detail. This includes the security implications of our design decisions and a number of relevant side-channels.
  
  \item We demonstrate two case studies: first, we present an end-to-end prototype based on Keystone~\cite{keystone} on an FPGA running a RISC-V processor~\cite{ariane} connected to multiple external peripherals (IO devices, sensors, etc.) emulated by an Arduino microcontroller. Our modifications to the software TCB of Keystone only amount to around 600 LoC. Second, we perform a case study based on a GPU-style accelerator~\cite{zaruba2020manticore} and integrate it within a \name{} while also supporting multi-tenant isolation. 

\end{enumerate}

\section{Background}
\label{sec:background}

\subsection{Keystone}
Keystone~\cite{keystone} is a TEE framework based on RISC-V similar to existing TEE designs such as Intel SGX~\cite{costan2016intel} and Sanctum~\cite{costan2016sanctum}. However, in contrast to these systems which leverage the MMU to isolate memory, Keystone isolates phyiscal memory using physical memory protection (PMP) to provide isolation. PMP is specified in the RISC-V privilege standard~\cite{riscv2019privspec} and its entries allow to configure access policies that can individually allow or deny reading, writing, and executing for a memory range. For instance, a PMP entry can be used to restrict the operating system (OS) from accessing the memory of the bootloader. Every access request to a prohibited range gets trapped precisely in the core and results in a hardware exception. In Keystone, the PMP entries are managed by the security monitor (SM) which runs in the highest privileged mode called m-mode. The untrusted OS runs in the supervisor mode (s-mode), whereas ordinary applications run in the least privileged user mode (u-mode). Isolated enclaves run in their own separate s and u-mode in parallel to the OS. 
The SM maintains its own memory separate from the OS and protected by a PMP entry. It facilitates all enclave calls, e.g., it creates, runs, and destroys enclaves. The SM configures the PMP entries so that the OS can no longer access the enclave's private memory. Upon a context switch, the SM re-configures the PMP to allow or block access to the enclave. For instance, during a context switch from an enclave to the OS, the SM changes the PMP configuration such that access to the enclave memory is prohibited. Conversely, on a context switch back to the enclave, the PMP gets reconfigured to allow accesses to enclave memory. 
Since the SM is critical for the security of any enclave and the whole system, it aims to be very minimal and lean. As such, the SM is orders of magnitudes smaller than hypervisors and operating systems (15k LoC vs millions LoC~\cite{torvalds2020linux,barham2003xen}). There are also efforts to create formal proofs for such a SM~\cite{lebedev2019sanctorum}. Keystone also provides extensions for cache side-channel protections using page coloring or dynamic enclave memory. 

\subsection{Device Tree}
The device tree is a list that accurately describes the physical memory mappings of a platform. It describes the central processor, i.e., its speed, its ISA, and at what address its cache starts. It also includes the DRAM base address and various other components on the die, such as various internal and external buses. It is usually used by the bootloader and the OS to bootstrap the system. As some peripherals cannot be detected automatically, they must be present in the device tree, as otherwise they will not get recognized by the OS. The device tree is usually burnt into ROM and available to the bootloader and the OS. It can therefore be considered trusted.

\section{Problem Statement}
\label{sec:problemStatement}

Modern platforms are composed of (disaggregated) heterogeneous devices, from simple sensors that measure temperature or humidity to complex accelerators for machine learning. We summarize all of these devices under the term \emph{\sphw} in this paper.
Many modern workloads are critically dependent on such \sphw and often handle sensitive data, e.g., patient records for machine learning. However, existing solutions contain severe limitations for such applications.

\subsection{Existing Solutions}

There are several existing solutions for applications that handle sensitive data while also leveraging \sphw. For example, a fully dedicated system could, of course, support such an application, but it would incur high costs and very poor flexibility. On the other hand, the application could be executed on an ordinary operating system or even in a virtual machine. However,  both of these approaches rely on substantial codebases with millions of lines of code~\cite{torvalds2020linux,barham2003xen} and likely contain a large number of lingering vulnerabilities. Finally, modern TEEs such as Intel SGX, RISC-V Keystone, and ARM TrustZone provide security guarantees to the applications while excluding the OS or the hypervisor. However, existing TEEs cannot easily be retrofitted to support such an application as they do not extend to \sphw. Intel SGX and Keystone, for example, rely on the untrusted OS to communicate with specialized hardware. On the other hand, ARM TrustZone provides isolated communication between enclaves and \sphw. Nevertheless, ARM TrustZone requires trusting the entire secure OS, including device drivers not used by the enclave.

\subsection{Alternative Approaches}
In this paper, we investigate an approach based on TEEs. However, approaches based on microkernels, such as seL4~\cite{klein2009sel4}, are also promising. We want to stress that both of these options would require significant changes, and both are bound to encounter challenges along the way. Microkernels have the advantage of already supporting applications that leverage specialized hardware, but, in turn, they do not support attestation.
TEEs, on the other hand, support many desired properties out of the box but lack integration with \sphw. The TCB of both approaches would probably be comparable with only a slight difference because microkernels include the scheduler in their TCB. Nevertheless, we believe both directions to be promising, but we focus on TEEs in this work. Further discussion on a potential approach based on a microkernel can be found in \Cref{sec:relatedWork}.

\subsection{Attacker Model}
\label{sec:problemStatement:attackerModel}

The attacker model is tightly coupled with the type of \sphw. We separate the \sphw into two classes due to their distinct effect on the attacker model: 

\begin{description}
\item [\Sphw with physical interaction:]
Such devices range from input-only, such as input peripherals (e.g., mouse, keyboard) and sensors (e.g., temperature sensor) to output-only devices (e.g., monitor) and combined IO devices (e.g., touchscreen). For any such device, a local physical adversary can manipulate the environment and thus the input (and potentially the output). E.g., a physical adversary can point a laser at a light sensor, thus changing the sensor's reading but not the room's overall light intensity. Hence, any \sphw that interacts with its physical environment cannot tolerate a physical adversary.

\item [\Sphw without physical interaction:]
There are \sphw units that do not explicitly interact with their environment. They draw power and produce heat, but their input and output are not related to the environment. GPUs and other accelerators are the prime examples of this class of \sphw, for whom a local physical adversary can be tolerated. 
\end{description}

In this paper, we assume a remote attacker who remotely controls the entire software stack, i.e., the OS and hypervisor. While the remote attacker model is a weaker assumption compared to the local physical attacker considered in the existing TEEs, the former covers both aforementioned classes of \sphw. In the remote model, the attacker cannot access the platform physically or hot-swap a \sphw device. Note that the untrusted OS is still in charge of managing \sphw, and thus is able to remap the devices or send a reset or power-off signal. In addition, an adversary may launch DMA attacks using rogue peripherals.
We assume that the CPU firmware is trusted. Similar to other TEE proposals, side-channel attacks are out of scope~\cite{costan2016intel}. However, we will discuss the implications of our proposal on existing side-channel attacks and defenses in \Cref{sec:securityAnalysis}. Finally, we consider denial-of-service attacks to be out of scope. 

\subsection{Security Goals}

\paragraph{G1: Enclave protection}
The enclave's private data must remain confidential and integrity protected at all times. This includes protection from malicious enclaves, DMA attacks, and rogue \sphw.

\paragraph{G2: Secure Integration with \sphw}
\Sphw must be able to be integrated into an enclave and their communication must remain confidential and integrity protected in all circumstances. 

\paragraph{G3: Attestation}
Attestation to an enclave should not only cover its code and the genuineness of the processor but also the involved \sphw.

\section{Overview}

\begin{figure}[tbp]
    \centering
    \includegraphics[width=0.7\linewidth]{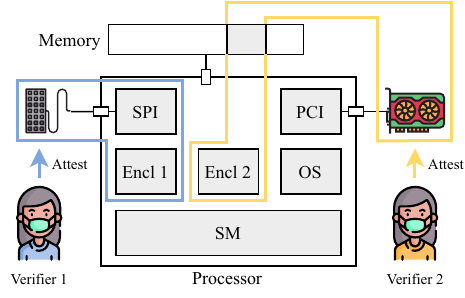}
    \caption{Two composite enclaves are highlighted by blue and yellow outlines. Both consist of two \nameenclave{}s: \texttt{Encl1} and the keyboard that is connected over the memory-mapped SPI bus, and \texttt{Encl2} and a GPU connected over PCI through DMA.}
    \label{fig:new_system}
\end{figure}

We propose a heterogeneous TEE architecture with a configurable hardware and software TCB. The enclaves that run on top of our design are called composite enclaves. As their name suggests, composite enclaves combine multiple components, such as a normal enclave on the CPU and a specialized hardware device such as an accelerator. To simplify, we call all these individual components \emph{unit enclaves}. In the following, we highlight how composite enclaves are constructed, starting with how individual unit enclaves communicate, what happens on a failure, and finally, how composite enclaves are attested.

\begin{figure}[tbp]
    \centering
    \includegraphics[width=0.4\linewidth]{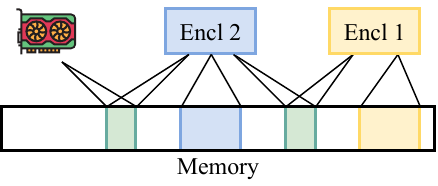}
    \caption{Example of private and shared memory regions with two enclaves, and a peripheral. Note that the shared memory region between the peripheral and Encl 2 can either be MMIO registers, and thus not backed by actual DRAM, or a DMA region.}
    \label{fig:memorymodel}
\end{figure}

In modern platforms, the processor communicates with specialized hardware devices using two mechanisms: memory-mapped IO (MMIO) or direct memory access (DMA). In our design, unit enclaves communicate over shared memory. We leverage existing memory protection mechanisms, such as PMP~\cite{riscv2019privspec} or TZASC~\cite{armtzasc400}, which already allow protecting any memory region, including MMIO and DMA regions. However, this implies sharing memory between enclaves, potentially endangering confidential data. We propose an architecture where every enclave has its own private memory and separate shared memory regions depicted in \Cref{fig:new_system}, and \Cref{fig:memorymodel}.

However, any of these communicating unit enclaves may encounter failures or other complications at any time, e.g., the unit enclave on the processor might get killed or destroyed without the keyboard noticing. In all of these edge cases, our proposal ensures that no confidential data is leaked (\textbf{G1} and \textbf{G2}). We achieve this by de-constructing all possible situations into two new enclave life cycle events: connect and disconnect. Intuitively, we provide a way to handle disconnects asynchronously by moving any shared memory region to the sole ownership of the surviving unit enclave. Follow-up synchronous disconnect and connect events may be employed to reestablish new shared memory regions and continue execution.

As mentioned before, our design must support an improved attestation mechanism that includes specialized hardware devices and the communication set up between the devices and the unit enclave on the processor (\textbf{G3}). To provide such an attestation mechanism, we propose a system where the verifier attests to all unit enclaves individually, receiving unique identifiers of connected unit enclaves, and then chains the reports together. 
However, chaining attestation reports could be vulnerable to timely manipulations in between two such attestations. 
We describe a mechanism that ensures safe attestation of composite enclaves in the presence of such manipulation attacks (c.f. \Cref{sec:security:attestation}). 

\section{Composite Enclaves}
\label{sec:approach}

In this section, we describe composite enclaves in detail. Composite enclaves combine \nameenclave{}s on the processor and on \sphw devices. First, we discuss the different types of unit enclaves and the necessary changes to \sphw to make them compatible. Then we introduce a shared memory model that allows \nameenclave{}s to communicate with each other and \sphw securely. Next, we discuss how the enclave life cycle changes given these modifications and how a remote verifier can attest to a \name{}. Finally, we provide a software design that makes it easier to adapt for software developers.

\subsection{\NAMEENCLAVE{}s within a Composite Enclave}
\label{sec:overview:enclaves}

A \name{} consists of multiple \nameenclave{}s that run on different hardware components and securely communicate with each other. A \name{} may span several \nameenclave{}s on the CPU and on \sphw. In the following, we describe the two main \nameenclave types.

\paragraph*{\Nameenclave{}s on the CPU}
\label{sec:overview:enclaves:processorEnclave}

\Nameenclave{}s on the CPU are similar to traditional enclaves, e.g., their runtime memory must be isolated from the OS and should only be accessible to the \nameenclave itself. To achieve that, we use physical memory protection (PMP) from the RISC-V privilege standard~\cite{riscv2019privspec} as introduced by Keystone.
We further differentiate two types of \nameenclave{}s on the CPU in our software design (\Cref{sec:programmingModel}): \app{}s and \ce{}s which encapsulate the application and driver logic respectively. 

\paragraph*{\Nameenclave{}s on \sphw}
\label{sec:overview:enclaves:peripheralEnclave}
Most \sphw runs some firmware or even some custom code (e.g., graphic shaders) which must be included in the TCB of a \name.
E.g., the GPU and its firmware in Figure~\ref{fig:new_system} is part of the yellow \name. Some \sphw may only be usable for a single tenant at a time, whereas others may support multi-tenancy for multiple \nameenclave{}s running simultaneously.
Since a remote verifier also wants to attest to \sphw devices, they must be modified to support attestation. However, we stress that these modifications  remain rather small (c.f. \Cref{sec:eval:accel}) and are discussed in several upcoming device attestation standards by the industry~\cite{SPDMattestation,jager2017rolling}. %

\subsection{Changes to \SPHW} A wide range of \sphw{} devices have unique behavior and integrate differently into \name{}s.  In this paper, we try to cover most devices but stress that some special cases require further analysis. We start with the simplest \sphw device we can imagine, a simple sensor, to one of the most complex, a sophisticated accelerator for a data center. Most other devices should fall in between these two examples and thus require modifications between these two extremes. 

\paragraph*{Simple sensors} A temperature sensor or other simple sensors only requires a minimal form of attestation to be integrated into \name{}s. Specifically, it must contain some key material to sign statements about itself. This is mandatory for (remote) attestation of a \name that includes an attestation report of such a sensor. We note that upcoming standards by the industry~\cite{SPDMattestation,pcie_measurement,jager2017rolling} already propose such attestation mechanisms for various \sphw ranging from simple sensors to accelerators. Any simple sensor that already supports such an attestation standard can be integrated into \name{}s without any hardware changes.

\paragraph*{Accelerators} On the other hand, accelerators tend to be very complex and may require more extensive modifications. Similar to simple sensors, they must support attestation (e.g., PCIe attestation~\cite{pcie_measurement}), but they may also require some form of multi-tenancy. Consider data-center applications, where multiple stakeholders want to move multiple compute-intensive tasks from the CPU to an accelerator. %
The individual tasks' data should remain confidential and isolated, not only on the CPU but also on the accelerator. Thus, such an accelerator requires multiple isolated and attestable domains -- in other words -- \nameenclave{}s that run on the \sphw.

\subsection{Communication with \SPHW}
\label{sec:approach:comm}
\label{sec:approach:sharedMemory}

To enable \nameenclave{}s on the CPU and \sphw to communicate securely, we make the observation that these devices generally communicate over mapped address regions: They either use an address range that is not reflected in DRAM, so-called memory-mapped-input-output registers (MMIO), or a shared DRAM region accessed via direct memory access (DMA). To maximize compatibility with existing drivers and \sphw, we chose not to change this behavior. Instead, we isolate the address regions that are used in this communication. Existing memory protection mechanisms like PMP already allow restricting access to a specific memory address region. They also allow restricting access to other address regions that are not in the DRAM range\footnote{E.g., DRAM could occupy the address range \texttt{0x8000000 - 0xF0000000}, whereas other \sphw such as UART could reside at \texttt{0x4000000 - 0x4001000}.}. Therefore, our proposal does not require any changes to the processor, as mechanisms such as PMP are already part of many standards~\cite{riscv2019privspec,armtzasc400}. Note that address regions used by \sphw are either i) static, i.e., hardcoded, in the form of a trusted device tree file, or ii) dynamic, i.e., configured at runtime by the SM. In our design, the SM always maintains a complete overview of all such regions and only allows a single \nameenclave on the CPU to access an address region of a \sphw device.

While we made the changes mentioned above to the SM to support \sphw with both MMIO and DMA, they also enable an alternative way for enclaves to communicate: shared memory. This reflects a major difference to traditional TEEs because most traditional enclaves can only communicate through the untrusted OS\footnote{Concurrent work~\cite{yu2020elasticlave} has also shown how shared memory can improve the performance of enclaves significantly.}.

\paragraph*{Polling and interrupts}
\sphw is synchronized with the processor with either polling or interrupts. Polling requires the CPU to check at a predetermined rate if new data is available from the \sphw, and thus, it is fully compatible with \name{}s. On the other hand, interrupts enable the \sphw to notify the CPU that new data is available with the processor's hardware support. Typically, the operating system registers interrupt handlers which get called when an interrupt occurs. In RISC-V, interrupts can be delegated from the highest privilege mode to lower ones by using either the \texttt{mret} instruction to forward individual interrupts or \texttt{mdeleg} for all interrupts of a specific type~\cite{riscv2019privspec}. So, in our design, the SM delegates relevant interrupts to the interrupt handler of a \nameenclave{} instead of the OS\footnote{In order to differentiate between interrupts, the SM includes a driver for the interrupt controller.}. 
Note that our prototype currently does not implement interrupt-based synchronization, and hence, we only evaluate polling-based synchronization.

\subsection{Enclave Life Cycle}
\label{sec:lifeycle}
\label{sec:overview:awareness}
The untrusted OS manages \sphw devices; hence, the OS could remap any device or send a reset signal. E.g., a GPU handing sensitive data could be shut down by the OS and remapped to a different GPU during runtime. In such a scenario, the \name{} should stop sending sensitive data to the GPU until the remote verifier re-attests the new GPU and its \nameenclave{}.

Traditional enclave's life cycle includes three distinct states: idle, running, and paused. E.g., the enclave is first created and starts in the idle state. Then the enclave transitions to the running state after a call from a user. Due to a timer interrupt by the OS scheduler, it is paused. It resumed again as soon as the scheduler yields back to the enclave. 

\paragraph*{Attaching \sphw}
Before going into the life cycle details, it is crucial to understand how \sphw is \emph{attached} to the platform and initialized.
There are two types of initialization procedures: statically compiled in the device tree or dynamically mapped by a bus controller. 
The device tree describes the specific address ranges and model numbers of all statically connected \sphw devices. It is usually stored in on-chip ROM and is provided to the OS by a zero-stage boot-loader, and thus, it can be considered trusted.
Dynamically mapped devices are mapped by a bus controller and a driver to a DMA region. In our proposal, the bus controller's driver, which sets up the DMA region, has to be trusted (but it could reside in its own \nameenclave{})

\paragraph*{Changes during runtime}
In all \nameenclave{}s, we introduce two additional life cycle events to describe what happens when a shared memory region is altered. These are \emph{connect} and \emph{disconnect} that are needed due to the asynchronous nature of \sphw, as a disconnect event could happen at any time.

The asynchronous disconnects are very critical as a composite enclave could end up continuing to use a memory region that is no longer protected due to a disconnect. Additionally, composite enclaves might want to provide graceful degradation and should not crash completely upon a disconnect. We solve both issues by splitting the disconnect event into an asynchronous disconnect and a synchronous disconnect. We consider both \nameenclave{}s or \sphw of a shared memory region to have shared ownership over that region. If one of the entities dies, the other entity gains the sole ownership of the memory region. As such, an asynchronous disconnect leads to the sole ownership of a previously shared memory region. In turn, the untrusted OS can issue a synchronous disconnect command to the SM to free the shared memory region and notify the \name and all its \nameenclave{}s of the disconnect. We mandate that before any connect command, the \nameenclave{} must first receive a synchronous disconnect. If this was not the case, an adversary could disconnect a benign \sphw device and reconnect a malicious one without the enclave noticing.

We illustrate the behavior of \name{}s using an example scenario. \emph{\Nameenclave{} 1} ($E_1$) connected to \emph{\nameenclave{} 2} ($E_2$), which, in turn, is  connected to a \sphw device ($HW$). We denote the shared memory regions as $S_{\{E_1, E_2\}}$, and $S_{\{E_1, HW\}}$that is shared among $E_1$ \& $E_2$, and $E_1$ \& $HW$ respectively.

\setcounter{para}{0}

\mypara{$E_1$ is killed} In such a situation, the specific shared memory region $S_{\{E_1, E_2\}}$ should be destroyed. To do that, the SM performs an asynchronous disconnect of $E_1$ for $S_{\{E_1, E_2\}}$ resulting in sole ownership of $S_{\{E_1, E_2\}}$ by $E_2$. Upon the following synchronous disconnect $S_{\{E_1, E_2\}}$ gets fully destroyed.
An application may require any sensitive data from $E_1$ that still remains on $HW$ to be cleared. In such a scenario, $E_2$ will tell $HW$ to clear this data on the following synchronous disconnect. %

\mypara{$E_2$ is killed} All shared memory regions associated with $E_2$ (this includes the shared memory regions with both $E_1$ and $HW$) are immediately modified by the SM during the asynchronous disconnect. They are now solely owned by $E_1$ and $HW$, respectively. Zeroing out $S_{\{E_2, HW\}}$ also implicitly notifies $HW$ that $E_2$ has died, forcing the \sphw to reset.
    
\mypara{$HW$ is killed/disconnected} In the asynchronous disconnect, the SM immediately modifies $S_{\{E_2, HW\}}$ to $S_{\{E_2\}}$. At some later point, the OS must issue a synchronous disconnect, which invalidates $S_{\{E_2\}}$. This also results in the destruction of $S_{\{E_1, E_2\}}$ in case $E_1$ accesses $HW$ through $E_2$. From then on $E_2$ is available to connect to a new $HW$ (after attestation).

\setcounter{para}{0}

\subsection{Attestation of a Composite Enclave}
\label{sec:approach:attestation}

We extend the existing notion of attestation from traditional enclaves to \name{}s that run on multiple \sphw devices within the platform. Traditionally, attestation ensures the current state of an enclave through a measurement of the code. The standard attestation report of a traditional enclave contains the measurements of the enclave and the low-level firmware (e.g., the security monitor in RISC-V keystone or $\mu$Code in SGX). Both of which are signed by the platform key (known as the device root key). In contrast, the attestation of a \name{} must also reflect all included \nameenclave{}s and corresponding \sphw devices. 
A potential attestation mechanism for a \name{} could be a lengthy report containing all the components' measurements, including the \sphw{} (similar to related device attestation standards~\cite{SPDMattestation,pcie_measurement,jager2017rolling}). 
Contrary to that, we provide the verifier with an option to decide which other \nameenclave{}s he wants to attest. When the verifier attests a specific  \nameenclave, a list of identifiers of all connected \nameenclave{}s is provided alongside the attestation report. These identifiers are assigned by the SM and can be used to specify which \nameenclave{} one wants to attest. A verifier can then chose to attest some or all the connected \nameenclave{}s from the list of identifiers.

\paragraph*{\Nameenclave{} identifiers} 
Upon creation of a new \nameenclave, the SM assigns a unique identifier to it. This identifier uniquely determines the \nameenclave{}s participating in a specific shared memory region. When the \nameenclave{} is killed, the identifier may be reused for other \nameenclave{}s (c.f. \Cref{sec:securityAnalysis}).

\begin{figure}[t]
  \centering
  \includegraphics[trim={0 5cm 15cm 0}, clip, width=0.6\linewidth]{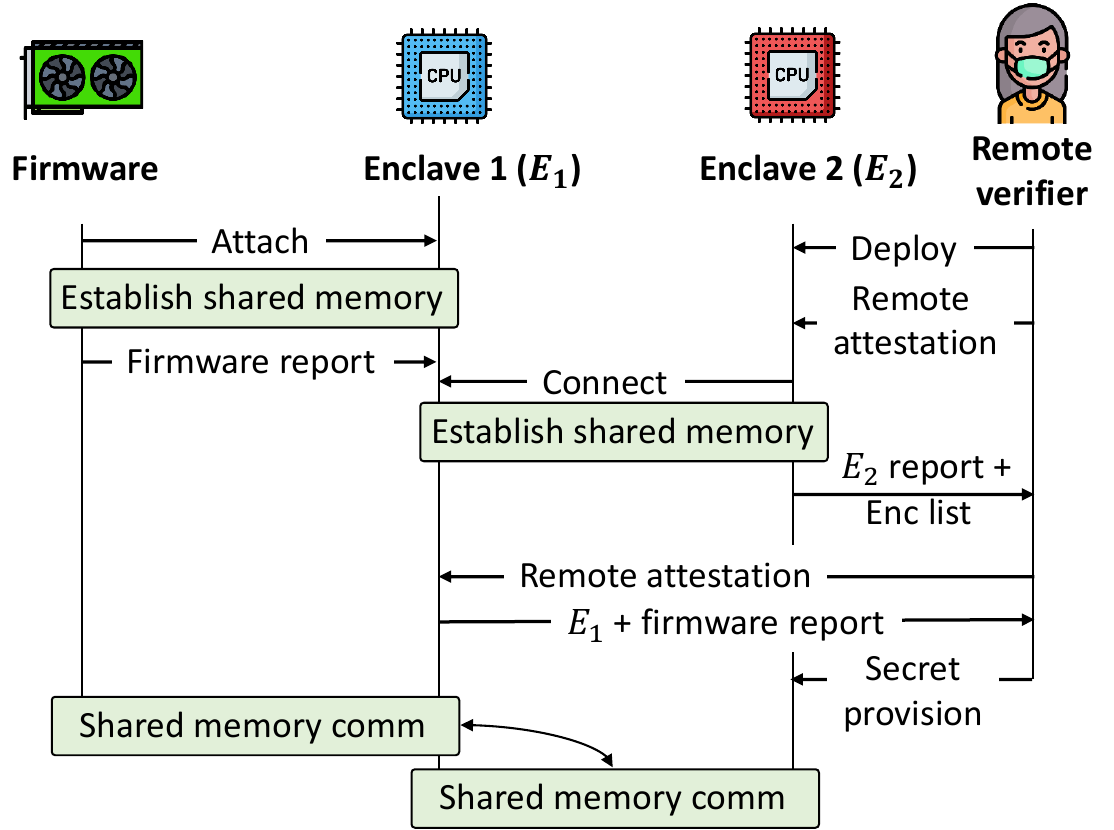}
  \caption{Flow of the remote attestation process between the user and a \name{} that consists of $E_1$, $E_2$ and the firmware.} %
  \label{fig:attestationFlow}
\end{figure}

\paragraph*{Attestation flow}

Figure~\ref{fig:attestationFlow} depicts an example \name{} and the sequence of the attestations between its different \nameenclave{}s. The example \name contains three \nameenclave{}s: \emph{enclave 1} ($E_1$), \emph{enclave 2} ($E_2$), and the firmware of a \sphw device. Note that the attestation process starts from the verifier who initiates a remote attestation request of $E_2$. 
The attestation report of $E_2$ includes a list  of connected \nameenclave{}s' identifiers, notably $E_1$. The verifier then executes a series of individual remote attestations to all connected \nameenclave{}s. Note that both individual attestations of $E_1$ and $E_2$ include each other's identifiers in their list of connected components. Also,  both the attestation reports of $E_1$ and $E_2$ are signed by the same platform key. This proves to the remote verifier that both \nameenclave{}s are running on the same platform.

For \sphw, the attestation mechanism is different. First of all, a \sphw device  needs to contain some key material and a signed certificate from the manufacturer. This allows a verifier to observe the legitimacy of the device. Secondly, the verifier from \Cref{fig:attestationFlow} needs to be able to verify that the \sphw is directly talking to $E_1$. This is facilitated by the SM, who checks the address regions for MMIO registers. DMA regions can even be established by an untrusted entity such as the OS. However, the attestation report of both the \sphw and $E_1$ contains the physical memory region that they share.

\subsection{Software Design}
\label{sec:programmingModel}

\begin{figure}[tbp]
     \centering
     \includegraphics[width=0.4\linewidth]{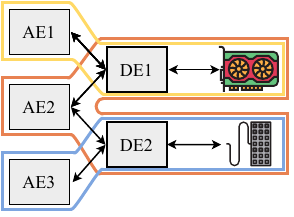}
     \caption{Three example \name{}s in our software design with \app{}s (AE), \ce{}s (DE), a GPU, and a keyboard. Note that the red \name{} is spanning over two external devices and \ce{}s isolate the data from the different \app{}s.}
     \label{fig:sharedMemory}
\end{figure}

In this section, we introduce \name{}'s software design which is one possible way for application, driver, and firmware developers to adapt their software to be compatible with \name{}s with minimal effort.

\subsubsection{Software components}
\label{sec:programmingModel:systemComponents}

Composite enclave's software design consists of three entities: \app{}s, \ce{}s, and firmware on \sphw devices, as shown in Figure~\ref{fig:sharedMemory}. \App{}s and \ce{}s are \nameenclave{}s on the CPU. \Sphw is connected to the platform over buses. Contrary to a monolithic design where the application and driver are fused into one big enclave, our modular approach aims to provide high flexibility and increase code reuse.

\setcounter{para}{0}

\myparagraph{\App{}s} \App{}s are similar to the traditional enclaves in Intel SGX or Keystone. In such TEEs, the enclaves cannot access \sphw without using the OS as a mediator, as the OS handles all drivers. \App{}s also cannot communicate with \sphw directly. The \app{}s use shared memory to communicate with a \ce that then communicates with the \sphw device. The rationale of separating the driver from the application logic is two-fold, i) to avoid requiring the developers to ship driver code with their application, and ii) one \ce per \sphw allows multiple \app{}s to communicate with that specific \sphw device in parallel. %

\myparagraph{\Ce{}s} The \ce contains the driver that facilitates communication with a \sphw device, and it may mediate any access to the device (e.g., rate-limiting). Note that \app{}s, standard non-enclave applications, and the OS can no longer access the \sphw device directly. The only way to communicate with the device is through the device-specific \ce. Such a design choice isolates the drivers: one compromised driver does not affect other \name{}s. The \ce maintains an isolated communication channel over shared memory to \app{}s and the \sphw device. To simplify the configuration, we assume that only one active \ce per \sphw exists at a time. However, any \ce can be replaced at the user's request. %

\subsubsection{Isolation of multi-\app session}\label{sec:programmingModel:systemComponents:multiApp} Multiple \app{}s could connect to a single \ce to have simultaneous access to a \sphw device. In such a scenario, the \ce keeps separate states corresponding to each of the \app{}s. Note that this is primarily a functional and then a security requirement as operations in one \app could affect the computation of another \app if there is no isolation. For some devices, the \ce may need to reset the state of the \sphw when it switches to a session with a different \app (temporal separation). However, sophisticated accelerators may support multiple isolated workloads in parallel (spatial isolation), and thus the state does not have to be reset.

\section{Security Analysis}
\label{sec:securityAnalysis}

In this section, we informally analyze the security of \name{}s. 
First, we show how isolation from a malicious OS (\textbf{G1}) and malicious \sphw (\textbf{G2}) is achieved including a number of relevant side-channel attacks. Then we analyze the life cycle events of \nameenclave{}s and discuss the security of the attestation of \name{}s (\textbf{G3}). 

\subsection{Isolation}

\paragraph*{Malicious OS}

We leverage PMP entries~\cite{riscv2019privspec} to protect address regions that are used by \nameenclave{}s. 
Recall that in stock keystone~\cite{keystone}, the PMP configuration only allows each enclave to access its private memory (\textbf{G1}). On top of this, we use additional PMP entries to protect shared memory regions (\textbf{G2}). 
Note that only the highest privilege level, i.e., the SM, can modify PMP entries. During a context switch, the SM re-configures all PMP entries such that the correct memory ranges are available again. The SM has the complete overview over all \nameenclave{}s and shared memory regions and sets up all PMP entries on its own. The processor will throw an access fault exception upon any memory access into protected memory regions. The hardware page table walker also must behave according to the configured PMP rules. Therefore, miss-configured page tables cannot be used to leak any data from protected memory ranges.

The SM enforces a shared memory region to be strictly shared between two entities (e.g., a \nameenclave{} on the CPU and a \sphw device). The SM also verifies that no overlap exists between the memory ranges similar to stock Keystone. 

\paragraph*{Rogue DMA requests}
Malicious peripherals may try to access protected memory through rogue DMA requests. However, mechanisms to mitigate such attacks already exist in most architectures, e.g., AMD IOMMU~\cite{amd2007iommu}, Intel VT-d~\cite{abramson2006vtd}, and ARM SMMU~\cite{arm2013smmu}. These mechanisms process every DMA request and verify its validity according to some access policy. Any memory access attempt that does not fit the access policy is blocked (\textbf{G1} and \textbf{G2}). Currently, there is no standardized mechanism to limit such DMA requests in RISC-V. However, there is a proposal of an input-output variant of PMP called IOPMP~\cite{IOPMP}. IOPMP enforces the configured PMP rules for non-RISC-V peripherals and mitigates DMA attacks completely. 

\paragraph*{Malicious application or \ce{}s}
The attacker-controlled OS can spawn malicious \app{}s and \ce{}s. However, users remotely attest before providing any secret to the \app. During the attestation, the user checks the attestation report of both the application and \ce and aborts if they do not match the intended enclave measurements. The attestation also reveals any misconfiguration of communication links by an adversary (\textbf{G2} and \textbf{G3}). Note that this only verifies the current configuration of communication links. Upon any change to this setup, the \app might require the external verifier to re-attest (c.f. \Cref{sec:security:lifecycleevents}). 

We require the \ce{} to provide isolation between multiple connected \app{}s (c.f. \Cref{sec:programmingModel:systemComponents:multiApp}). Hence an attacker-controlled \app cannot access the confidential data of other \app{}s in the same \ce{}.

Vulnerabilities within any of these \nameenclave{}s could break the isolation guarantees of the data in that specific \nameenclave{}. However, such an attack remains contained in the compromised \nameenclave{} and cannot spread to other connected enclaves. E.g., if a vulnerability in a \ce{} is found, only the data within that enclave is revealed. Any data that does not pass through the compromised \ce remains confidential. In this way, we provide defense-in-depth and reduce the potential impact of vulnerabilities.

\paragraph*{Malicious \sphw}
If an adversary manages to compromise the exact device used by a \name, then any data on the device is forfeit. However, any data not passed to the malicious device remains confidential (\textbf{G2}).
We stress that certain manipulations of specific peripherals are always possible for an adversary. Consider, for example, a temperature sensor. Any local physical adversary can increase the real-world temperature and thus manipulate the sensor reading. However, as we describe in our attacker model in Section~\ref{sec:problemStatement:attackerModel}, the physical attacker is out-of-scope of this paper. %

\paragraph*{Remapping Attacks}
Many \sphw devices are plug-and-play and thus dynamically mapped by the OS. Therefore, the OS may also change the mapping during runtime, potentially leading to confidential data being shared with the wrong entity. We analyze all types of dynamically mapped \sphw and how our proposal prevents such a remapping attack (\textbf{G2}).

Dynamically mapped \sphw devices can use one out of the following mechanisms: i) a DMA region which facilitates all communication, ii) a bus controller driver facilitates the communication, or iii) a mix of both of these. Note that MMIO interfaces are generally not dynamic and do not change during runtime. 

In remapping attacks against pure DMA devices, the OS may remap the DMA buffer to a different address range. There are two weak points where confidential data could leak: the unit enclave on the CPU could share confidential data with a remapped untrusted device, or the device could share results with the wrong entity on the processor. However, the OS needs to notify the device of the remapping (if this does not happen, the device will write to the wrong address), so the second potential leakage is ruled out immediately. In the other case, it is essential to note that the shared memory region of the unit enclave remains protected by PMP entries. 
Thus, even after remapping, the OS cannot access the shared memory region containing confidential data.

If the communication is (partially) facilitated by the bus controller, the bus controller and its accompanying driver must be part of the TCB since both of them process all communication and may leak confidential data. 

\paragraph*{Side-channel attacks}
While we do not evaluate any defenses against side-channel attacks, we discuss potential side-channel attacks against our proposal and how they could be mitigated. Many parts of \name{}s remain the same as in traditional TEEs, where side-channels have been widely investigated~\cite{brasser2019dr,brasser2017software,gruss2017strongsidechannel} (\textbf{G1}). However, we note that our approach creates some new side-channels that may not be present in traditional TEEs, such as bus contention (typically related to \textbf{G2}). 

Microarchitectural side-channels in traditional TEEs leverage shared resources such as the cache~\cite{brasser2017software}, branch predictor~\cite{lee2017inferring}, and memory translation~\cite{xu2015controlled}. There exist several defenses against such attacks. Spatial partitioning of the cache in the form of cache coloring can fully defend against all cache-based side-channel attacks~\cite{costan2016sanctum,zhang2009cachecoloring,zhaosonicboom}. Similarly, other proposals have called for cache randomization~\cite{brasser2019dr,werner2019scattercache}. Processor features such as transactional memory have also been shown to mitigate cache attacks with low overhead~\cite{gruss2017strongsidechannel}. To the best of our knowledge, all of these proposals can be applied to \name{}s due to the similar internal structure to traditional TEEs.
\Sphw contain shared resources such as caches, and thus are equally vulnerable as the processors~\cite{naghibijouybari2018rendered,visor,ramesh2018fpga}. However, mitigating these attacks is an orthogonal problem.

The introduction of \sphw into TEEs also implicates the bus as a new shared resource. An adversary could measure the throughput of her connection over the bus and observe any contention on the bus leading to less throughput.
Bus contention, however only exposes the bus access patterns. In extreme cases, the timing of bus contention could leak data, e.g., one side of the branch performs bus accesses while the other does not~\cite{paccagnella2021bus}. This behavior is very similar to previous timing and side-channel attacks, and there exist multiple mitigations, such as oblivious execution~\cite{rane2015raccoon}, that can be applied in the same way to the bus side-channel.

\subsection{Life Cycle Events}
\label{sec:security:lifecycleevents}
As described in \Cref{sec:lifeycle}, we introduce two additional life cycle events for \nameenclave{}s. \texttt{Connect} is used to connect two \nameenclave{}s over a shared buffer, whereas \texttt{disconnect} facilitates a disconnect. The \texttt{disconnect} is split into a synchronous and a asynchronous event. The asynchronous disconnect only occurs when one of the \nameenclave{}s unexpectedly dies and results in the transfer of the sole ownership of the memory region to the remaining enclave. This enclave can then try to continue its execution. However, it will realize that the other \nameenclave{} has died as it does not react to any activity on the shared memory region. At a later point, the untrusted OS can issue a synchronous disconnect to notify the \nameenclave{} and free the shared memory officially. Note that the SM mandates a synchronous disconnect before another \texttt{connect} command. Due to this architecture, a stale shared buffer will never be made accessible to any untrusted entity until a synchronous disconnect occurs, during which the \nameenclave{} will officially get notified. The separate handling of synchronous and asynchronous disconnect events enforces protection for any secret data during an enclave's entire life cycle (\textbf{G2}).

\subsection{Attestation} 
\label{sec:security:attestation}
As specified in \textbf{G3}, the attestation of a \name{} should also cover all connected \nameenclave{}s. In our proposal, the attestation report of a \nameenclave{} contains identifiers of all the connected \nameenclave{}s. The SM generates these identifiers and makes sure that no two running \nameenclave{}s share same identifier. Hence, a \nameenclave{} could be assigned with an identifier that belonged to a \nameenclave{} in the past. Of course, strictly increasing identifiers implemented with monotonic counters could be used for the identifier but such a solution needs a non-volatile storage on the CPU that might be expensive. 

Now assume that the adversary kills an \nameenclave{} and launches another \nameenclave{} with a \emph{different} binary (defined as \texttt{code}), but with the exact same identifier. I.e., the attacker can kill \nameenclave{} $A$ and launch $A'$, \texttt{code($A')\neq$code($A$)}, with the same identifier, \texttt{ID($A$)$=$ID($A'$)}. However, when a remote verifier attests $A'$, the verifier sees that the measurements mismatch as \texttt{code($A')\neq$code($A$)} and rejects it.

Lets assume a more complex scenario with two pairs of \nameenclave{}s: $A, B$ and $A', B'$, where $\texttt{code} (A')\neq \texttt{code} (A)$ but $\texttt{code} (B') = \texttt{code} (B)$. 
A remote verifier attests to a \nameenclave{} $A$ that is connected to $B$ and and establishes a shared secret with $A$. Before the verifier attests to $B$, the attacker kills $B$. The attacker then spawns a new \nameenclave{} $B'$ where \texttt{ID($B$)$=$ID($B'$)}. The remote verifier will then attest to $B'$ and find that the code measurement looks fine. However, we stress that $B'$ cannot be connected to $A$ because then $A$ would need to receive a synchronous disconnect and would need to be re-attested (due to the configuration of $A$). If the attacker also kills $A$ and replaces it with $A'$ (where \texttt{ID($A$)$=$ID($A'$)}) and connects $A'$ and $B'$. The verifier would then see that $B'$ has the correct measurement and is connected to the identifier of $A$ (as \texttt{ID($A$)$=$ID($A'$)}). However, the verifier will want to provide its data to $A$ using the shared secret they have established in the previous attestation. Obviously, this cannot succeed as the new \nameenclave{} $A'$ cannot know the secret.

\section{Implementation}
\label{sec:eval}

\begin{figure}[t]
\centering
\includegraphics[trim={0 8cm 19cm 0}, clip, width=0.6\linewidth]{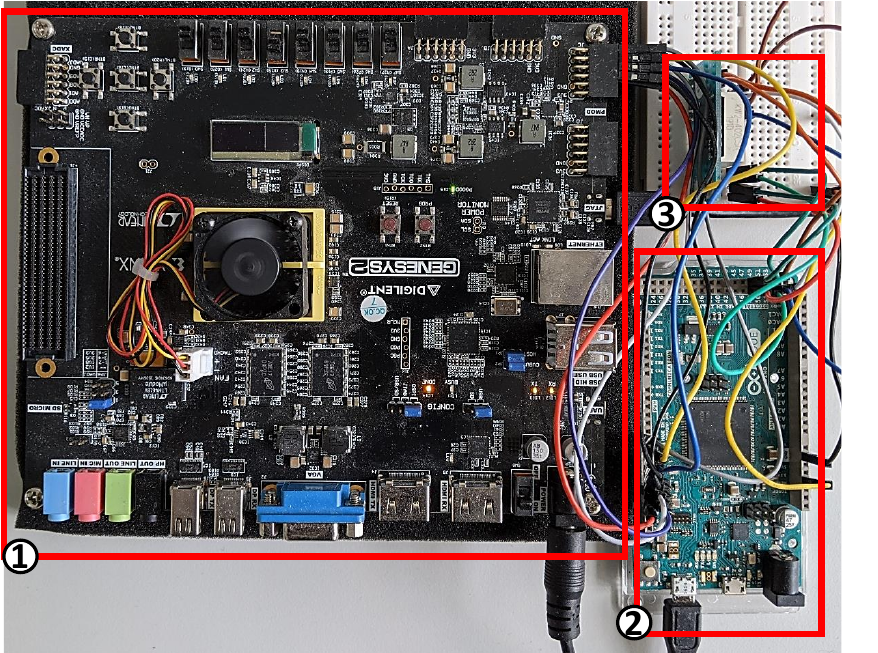}
\caption{Our prototype: \one Digilent Genesys 2 FPGA board, \two Arduino Due as the peripheral simulator, and \three a seven-segment display unit as an example peripheral.} %
\label{fig:prototype}
\end{figure}

\subsection{FPGA Prototype}
\label{sec:eval:prototype}
We implemented an end-to-end prototype of a \name on a softcore on an FPGA running a modified Keystone enclave framework~\cite{keystone} (available online~\cite{ourcode}). Figure~\ref{fig:prototype} shows one of our experimental setups consisting of an FPGA emulating the central processor connected to several Arduino boards that emulate \sphw. %

\paragraph*{FPGA platform}
We base hardware platform on the Ariane core~\cite{ariane}, an open-source RISC-V 64-bit core that supports commodity OS such as Linux. It is an RC64GC 6-stage application class core that has been taped out multiple times and can operate up to 1.5 GHz. We run this core on a Digilent Genesys 2 FPGA board (\one in~\Cref{fig:prototype}).

Since the core originally did not support PMP, we added PMP capability in around 160 lines of SystemVerilog. The PMP unit is formally verified with a bounded model check against a handwritten specification with yosys~\cite{wolf2016yosys}. Two of these units are inserted into the memory management unit (MMU) and are responsible for checking data accesses and instruction fetches. An additional unit is placed in the hardware page table walker to check page table accesses.
Our implementation has a configurable number of PMP entries up to the maximum number of 16 mandated by the standard~\cite{riscv2019privspec}. Our modifications have been contributed to the Ariane project and are open source~\cite{arianegithub}. Note that PMP is part of the RISC-V privilege standard and as such is already available on many other cores~\cite{asanovic2016rocket,ibex}. 

\paragraph*{Modifications to Keystone}
We modified the SM to be able to connect two \nameenclave{} or an \nameenclave{} and \sphw. Specifically, we added three new interfaces to the SM called \texttt{connect}, \texttt{sync\_disconnect}, and \texttt{async\_disconnect}. These interfaces can be used to set up shared regions between two  \nameenclave{}s or \sphw specified by their identifier. We also modified Keystone's attestation procedure to include a list of identifiers for all connected  \nameenclave{}. Our modifications only amount to 390 additional or modified lines of code. The SM consists of around 2000 lines of code excluding SHA3 and ed25519 implementations that contribute around 4000 additional lines of code. %

Every enclave runs on top of a trusted minimal runtime that handles syscalls and manages virtual memory. For our prototype, we added support to dynamically map shared memory regions into the virtual address space of a \nameenclave{}. We modified 213 LoC out of 3600 LoC for Keystone's runtime.

\paragraph*{Simple \sphw} 
In our prototype, we emulate a number of simple \sphw (e.g., keyboard, mice, simple sensors, etc.) on the Arduino Due microcontroller prototyping board (\two in Figure~\ref{fig:prototype}) using the Arduino HID library. The Due's GPIO pins are connected to the FPGA's PMOD pins over two pairs of $8$ wires for bi-directioanl data. We modify the $I^2C$ protocol to communicate data between the Due and the FPGA. The physical limitations of the PMOD pins restricts the channel's frequency to $8$ MHz yielding 1 MB/s bandwidth. In the real world, the physical interfaces between the \sphw and the platform could be diverse such as USB and PCI-E. As a concrete example, we implemented a keyboard with the Arduino board and wrote a simple keyboard driver that interprets the GPIO signal from the Arduino. Additionally, we use a PMOD interface-based seven-segment display unit as an output peripheral (\three in Figure~\ref{fig:prototype}). The driver contains around 50 LoC and is incorporated into our example \ce. Additionally, we use the \texttt{USBHost} library that can emulate a number of USB peripheral devices on the Arduino. We use the Arduino cryptographic library for signing the challenge messages from the \ce during the local attestation. The Due uses 128-bit AES (CTR mode) for encryption, HMAC\_SHA256 for message authentication, Curve25519 for key exchange, and SHA3 for the hash function. We use \texttt{DueFlashStorage} library to implement the NVM flash that contains the key material for the peripheral attestation. Our prototype implementation is approximately 2.5K lines of code.

\begin{figure}[tbp]
\centering
\includegraphics[width=0.5\linewidth]{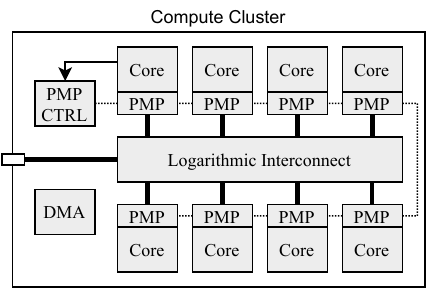}
\caption{An architecture overview of one compute cluster of our modified accelerator with one PMP control unit per cluster and individual PMP enforcement units per core.} %
\label{fig:accelerator_arch}
\end{figure}

\subsection{Accelerator}
\label{sec:eval:accel}

We conduct another case study to show how complex \sphw such as a GPU-scale accelerator~\cite{zaruba2020manticore} can be extended to support \name{}s. The accelerator is a 4096-core RISC-V platform that has comparable performance to current machine learning accelerators. It is organized in clusters each with 8 individual single-stage RISC-V cores~\cite{zaruba2020snitch}, each of which is accompanied by a double precision floating point unit capable of two double precision and four single precision flops per cycle. To hide memory latency, all clusters have access to a scratchpad memory and a large L2 data cache. 

To provide multi-tenant isolation on the accelerator, we introduce a shared PMP control unit with 4 entries into every cluster. Every core then has its own PMP enforcement unit. The PMP entries can only be configured by one out of eight cores but the access policies will be enforced on all of them. 
The architecture of the modified compute cluster is shown in \Cref{fig:accelerator_arch}. 
With this additional hardware support we were able to implement a small firmware that configures the PMP entries according to the specifications from the host and then runs a task in user mode. Upon a context switch, the scratchpad memory that was in use by the previous task is flushed and the PMP entries are reconfigured. The firmware consists of 143 lines of assembly and 73 lines of C code. %

\section{Evaluation}
\label{sec:eval:numbers}

\paragraph*{Performance of Inter-Enclave Communication}

As \name{}s supports shared memory to communicate, its communication speed is the same as what the memory bus provides. This is much faster compared to traditional TEEs, where enclaves communicate through the OS requiring extra encryption steps. Hence, we do not believe a comparison between these two systems is meaningful. Concurrent work also demonstrates the performance gains by using shared memory between enclaves~\cite{yu2020elasticlave}.

\begin{figure}[t]
    \centering
    \includegraphics[width=0.7\linewidth]{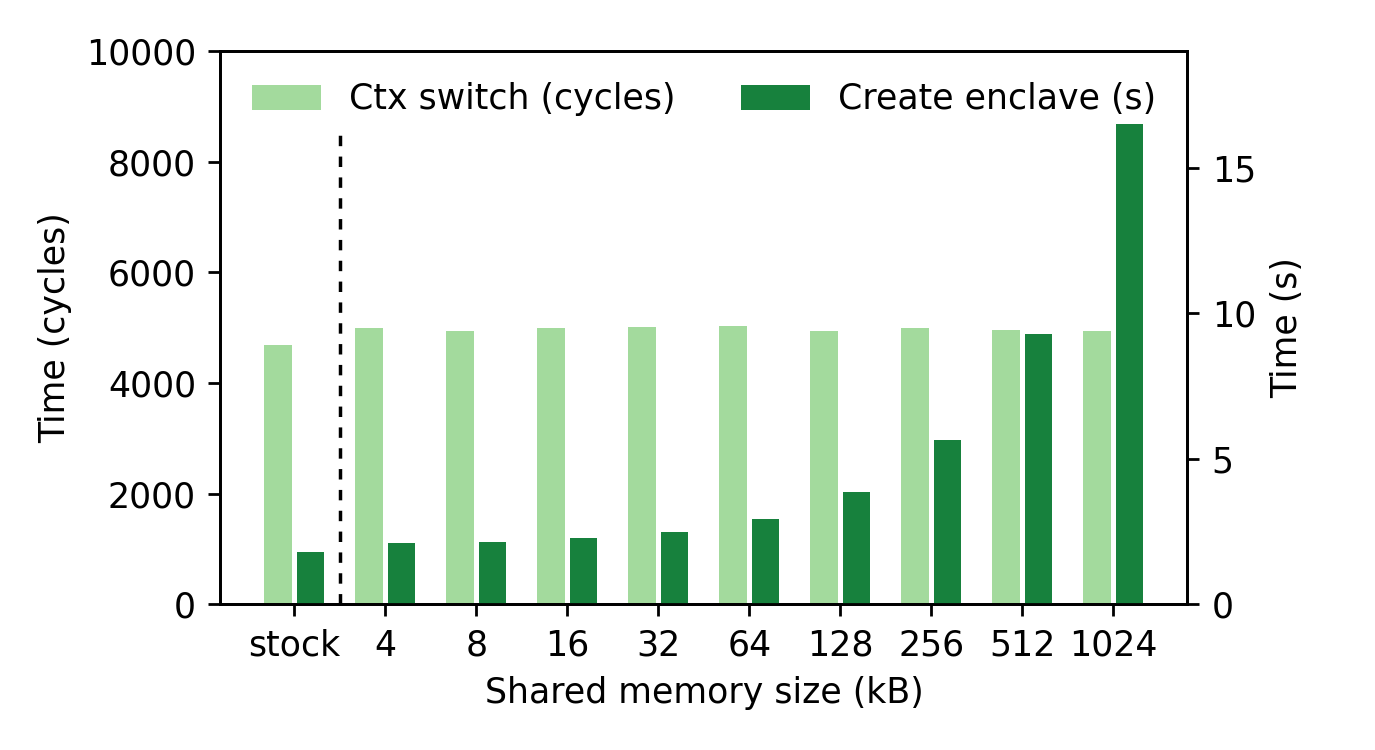}
    \caption{Context switch performance for varying sizes of a shared memory region compared to stock Keystone performance on the left (equivalent to no shared memory).}
    \label{fig:ctxswitches}
\end{figure}

\paragraph*{Context Switch Performance}
Context switches are critical for any system and determine its responsiveness and a part of its performance.
We performed experiments for various sizes of shared memory region and gathered various context switch latencies in \Cref{fig:ctxswitches}. We also measured the time of \nameenclave{} creation which is mostly dominated by copying all the \nameenclave{} data from the untrusted OS to the protected memory region and thus is expected to be linear in terms of memory size. Our measurements highlight that the context switches are independent on the shared memory size. The absolute context switch time increases from 4730 for stock Keystone to 4950 for our prototype.

\paragraph*{PMP Overhead}
We measure the hardware overhead of PMP units in terms of the logic, the caches, and the total amount in NAND2 gate equivalents within the Ariane processor pipeline for 0, 8, and 16 PMP entries in \Cref{tab:eval:ariane}. We instantiate the Ariane core~\cite{ariane} with the default configuration: including the floating point unit, 32KiB L1 data cache, 16KiB L1 instruction cache, branch history table of size 64, and a 16-entry branch target buffer. We synthesized this core configuration in a 22nm technology at 1GHz. %

\begin{table}[tbp]
    \centering
    \caption{Size of the default configuration of the Ariane core in gate equivalents (GE), synthesized in 22nm at 1GHz with varying number of PMP entries.}
    \begin{tabular}{llll}\toprule
        PMP entries & logic & caches & total \\\midrule
        0 & 472k GE & 686k GE & 1141k GE \\
        8 & 497k GE & 686k GE & 1164k GE \\
        16 & 531k GE & 686k GE & 1197k GE \\ \bottomrule
    \end{tabular}
    
    \label{tab:eval:ariane}
\end{table}

\paragraph*{IO Peripherals} The communication overhead between the platform and the peripheral device emulated by the Arduino due is very small. At the time of initialization, the peripheral and the platform exchanges handshake messages to perform local attestation. The initial handshake message is $60$ bytes. Every message size of our modified $I^2C$ protocol is 32 bytes. The combined latency introduced by signing averages around 60 $\mu$s.

\paragraph*{Accelerator}
Our modification of the accelerator cores slows down from 750MHz to 666MHz due to the impact of the PMP access checks on the critical path. Note that this may not reflect the general case. The change in area of a single core complex (core, FPU, and an integer subsystem) can be found in \Cref{tab:areasnitch} for 750 and 666 MHz respectively. Note that the size of the core increases due to the increased pressure by the PMP, while the FPU and the IPU get smaller with the lower clock as their critical path is not affected by the PMP entries. In total, the area of the entire accelerator decreased by around 0.7\% while the clock frequency was reduced by 15\%.

\begin{table}[tbp]
\centering
\caption{The area rundown of the accelerator with 0 and 4 PMP entries respectively. Synthesized in 22nm with 750MHz clock for 0 entries and 666MHz with 4 entries.}
\label{tab:areasnitch}
\begin{tabular}{@{}lccc@{}}
\toprule
\multirow{2}{*}{Area {[}$\mu$m$^2${]}} & \multicolumn{2}{c}{PMP Entries} & \multirow{2}{*}{Overhead} \\ & 0                 & 4                 &\\ \midrule
Core   & 5.7               & 6.7               & 15.5\%                    \\
FPU    & 39.2              & 37.9              & -3.3\%                    \\
IPU   & 8.6               & 8.5               & -1.4\%                    \\ \midrule
Total  & 53.5              & 53.2              & -0.7\%                    \\ \bottomrule
\end{tabular}
\end{table}

\setcounter{para}{0}

\section{Limitations}
\label{sec:limitation}

\paragraph{Remote Attacker Model} 
As mentioned in \Cref{sec:problemStatement:attackerModel}, we only consider a remote adversary throughout this paper. For some use-cases it is impossible to consider a local phyiscal adversary who could, for example, change the environment that is measured via a sensor. However, this fundamental limit does not apply to more sophisticated \sphw such as accelerators. In this case, our proposal could be extended by two hardware modifications to cope with a phyiscal adversary: First, the CPU needs to support memory encryption and integrity, a typical mechanism that many TEEs already employ~\cite{sgx,kaplan2016amdmemencryption,suh2003aegis}. Second, the communication channel between \sphw and the CPU, i.e., the bus, must provide confidentiality and integrity. Existing proposals from industry and academia~\cite{gueron2016memoryencrpytion,kaplan2016amdmemencryption,suh2003aegis} indicate that such encryption capabilities are feasible and might become available in the near future.

\paragraph{Limited Number of PMP Entries} The number of PMP entries in the RISC-V privilege specification is limited to 16 (an extension to 64 is in discussion). This limits the number of \nameenclave{}s and shared memory regions that may coexist on a system. Assuming one shared memory region per \nameenclave{}, at most $(N-2)/2$ \nameenclave{}s can exist at a time (16 entries support 7 \nameenclave{}s). However,  isolation of \nameenclave{}s could also be achieved using the memory management unit (MMU) in a similar fashion as Intel SGX~\cite{costan2016intel}.
MMU-based isolation can also easily be extended to shared memory ranges and remove any limitation on the maximum number of \nameenclave{}s. 

\paragraph{Large Drivers}

\Sphw devices can be very complex and require major drivers to work. As an example, an open-source driver for AMD GPUs in the Linux kernel occupies around 3.3 million LoC~\cite{torvalds2020linux} (most of it are generated header files, 500k LoC without headers). Moreover, such drivers also leverage other capabilities of the kernel, and moving such a driver into a single \ce{} would require to replicate these capabilities. However, such a driver (e.g., for a GPU) was not created for a minimal TCB but for feature completeness. It could be possible to strip such drivers to the bare minimum needed to support the actual \app{}.

\section{Related Work}
\label{sec:relatedWork}

\paragraph*{TEE-based Solutions} There exist a number of solutions for integrating external devices into TEEs. SGXIO~\cite{weiser2017sgxio} aims to allow Intel SGX enclaves to interact with IO devices under the remote adversary model. SGXIO uses a trusted hypervisor which virtualizes peripherals. However, SGXIO is static, i.e., all the peripherals have to be set up at boot time and no changes are allowed during runtime (connect new peripherals, etc.). %

There are various proposals that aim to extend TEEs to GPUs~\cite{jang2019HIX,volos2018graviton}. While some only allow using the entire GPU in an enclave~\cite{jang2019HIX}, others also enable multi-tenant usage of GPUs~\cite{volos2018graviton}. Such multi-tenant GPU TEEs would fit very well within a \name{} as it is an excellent example of an enclave on \sphw and it shows that even some of the most powerful accelerators can be extended with a local TEE. Visor~\cite{visor} goes even further and proposes a hybrid TEE that spans over both CPU and GPU and their communication. Visor is aimed towards privacy-preserving video analytics where the computation pipeline is shared between the CPU (non-CNN workloads) and the GPU (CNN workloads) to increase efficiency. HETEE~\cite{zhu2020hetee} is another proposal to extend TEEs to GPUs without requiring changes to existing hardware. HETEE focuses on datacenter applications and proposes an extra hardware box per rack that is protected from physical attacks and contains all GPUs. Each enclave then runs on a dedicated compute server and a connected accelerator. In essence, the HETEE box provides secure routing of accelerators to dedicated compute servers. In contrast to HETEE, we aim to be able to execute multiple composite enclaves on the same machine. 

ARM TrustZone is a system TEE provided by ARM for their system-on-chips~\cite{winter2008trusted}. TrustZone applications run on top of a secure OS that is trusted and isolated from the standard OS (rich OS). 
TrustZone only provides the lower level isolation property between the rich OS and the secure OS with an extra bit on the bus. Everything else, i.e., isolation between TrustZone applications or remote attestation, has to be implemented by the secure OS~\cite{ning2014samsungknox}. Due to this limitation, manufacturers usually only allow TrustZone applications that are signed by them. 
Sanctuary~\cite{brasser2019sanctuary} extends TrustZone with user-space enclaves. Sanctuary achieves isolation by running enclaves in their own address space in the normal world. However, it does not extend to external \sphw. Some other proposals~\cite{TruZ-Droid,trustUI,SeCloak,VButton} enable additional security properties such as a trusted path by enabling direct pairing of peripherals (e.g., the touchscreen) to TrustZone applications. However, these are only geared towards IO for trusted path and do not support generic (dynamic) devices.

Finally, CURE~\cite{bahmani2021cure} proposes a TEE architecture that enables enclaves on all privilege levels. As such, CURE also enables enclaves that have exclusive access to specific peripherals against a software adversary similar to our approach. However, attestation to an enclave in CURE does not extend to peripherals. Besides, kernel-space enclaves in CURE run on a reserved core with, to the best of our knowledge, no option to yield back to the OS, and thus,  wasting resources while waiting for new data from peripherals.

\paragraph*{Other Isolation Methods} Minimal hypervisors or microkernels~\cite{herder2006minix} can also achieve isolation, and some are even formally verified~\cite{klein2009sel4,vasudevan2013design}. Usually, such proposals do not natively support attestation. However, by adding a root-of-trust and some minor software components to measure and sign applications, microkernels could be extended to support a simple form of remote attestation similar to academic TEE proposals~\cite{keystone}. While there are other challenges to overcome such as key distribution and revocation, and software updates, these challenges are identical for TEEs and have been handled in the past~\cite{sgx,kaplan2016amdmemencryption}.
From a TCB perspective, hypervisors and microkernels include a scheduler, moving it from the untrusted OS to the TCB, and thus, may result in a bigger TCB.

\paragraph*{Bump in the Wire-based Solutions} Fidelius~\cite{fidelius}, ProtectIOn~\cite{protection}, IntegriScreen~\cite{integriscreen}, FPGA-based overlays~\cite{fpga_overlay}, IntegriKey~\cite{integrikey} are some of the trusted path solutions that use external trusted hardware devices as intermediaries between the platform and IO devices. These external devices create a trusted path between a remote user and the peripheral and enable the user to exchange sensitive data securely with the peripheral in the presence of an attacker-controlled OS. %

\paragraph*{Related Standards}

Recently, there have been multiple upcoming standards backed by major players from the industry focused on new bus architectures~\cite{CCIX,CLX}. These proposals are motivated by the move to more \sphw and to disaggregated computing. CCIX~\cite{CCIX} tries to extend PCIe with a cache coherency protocol to allow multiple chips to have the same view of memory. All chips connected with CCIX may have their own memory, cache, and compute. However, all chips interconnected with CCIX are equally privileged, leading to a rather bleak security outlook for CCIX.

The other upcoming standard, CLX~\cite{CLX}, assumes current platform architecture similar of today with a host processor connected to multiple accelerators. As such, CLX is able to simplify the protocol by following a master-slave principle. CLX allows accelerators to cache shared memory. As such, the interaction between the CPU and accelerators no longer need expensive copying operations and both may even operate on the same data at the same time. CLX also has some provisions for link-encryption leveraging authenticated encryption to defend against bus tapping attacks. However, CLX is only a bus architecture and does not consider and adversary in either the accelerator or the host. Nevertheless, CLX would be a prime candidate to integrate into \name{}s.

\setcounter{para}{0}

\section{Conclusion}
\label{sec:conclusion}

We introduce \name{}s, a disaggregated TEE with a configurable hardware and software TCB. Composite enclaves allow to integrate \sphw into TEEs, something that before was not easily possible without concessions for more software TCB.
We present a prototype based on RISC-V and two case studies: i) emulated peripherals connected to an FPGA board, and ii) a GPU-style accelerator. Our evaluation of both case studies demonstrates low performance-overhead for context switches and feasibility for a wide range of \sphw devices.

\section*{Acknowledgments}
We thank the anonymous reviewers for their valuable feedback, and Petr Svenda for the seamless shepherding process. Thanks to Kaveh Razavi and Shwetha Shinde for feedback on early versions of this paper.

\bibliographystyle{alpha}
\bibliography{bibliography}

\end{document}